\documentclass[pdflatex]{ws-ijmpcs}
\usepackage{amssymb,amsmath,epsfig}
\def\trimmarks{}
\DeclareMathOperator{\tr}{Tr}
\newcommand{\sT}{{\scriptscriptstyle T}}

\renewcommand{\d}{\mathrm{d}}
\newcommand{\qT}{\bm{q}_\sT}
\newcommand{\mc}[1]{\mathcal{#1}}

\begin{document}

\markboth{Cristian Pisano}
{}

%
\catchline{}{}{}{}{}
%

\title{Transverse momentum dependent gluon distributions at the LHC}

\author{Cristian Pisano}

\address{Nikhef and Department of Physics and Astronomy, VU University Amsterdam, De Boelelaan 1081\\
NL-1081 HV Amsterdam, The Netherlands\\
cristian.pisano@nikhef.nl}

\maketitle

\begin{history}
\received{Day Month Year}
\revised{Day Month Year}
\published{Day Month Year}
\end{history}

\begin{abstract}

Linearly polarized gluons inside an unpolarized proton contribute 
to the transverse momentum distributions of (pseudo)scalar particles produced in hadronic collisions, such as Higgs bosons and quarkonia with even charge 
conjugation ($\eta_c$, $\eta_b$, $\chi_{c0}$, $\chi_{b0}$). Moreover, they 
can produce azimuthal asymmetries in the associated production of a photon and a $J/\psi$ or a $\Upsilon$ particle, in a kinematic configuration in which they are almost back to back. These observables, which can be measured in the running experiments at the LHC, could lead to a first extraction of both the polarized and the unpolarized gluon distributions and allow for a study of their process and energy scale dependences.

\end{abstract}

\ccode{PACS numbers:12.38.-t; 13.85.Ni; 13.88.+e}

\section{Introduction}

The momentum distribution $g(x)$ of unpolarized gluons, with collinear momentum fraction $x$ inside an unpolarized proton, has been investigated widely, both theoretically and experimentally. Its knowledge is of fundamental importance because it underlies most of the reactions which are currently being investigated at the LHC. Within the framework transverse momentum dependent (TMD) factorization, this distribution is a function of the gluon transverse momentum $p_\sT$ as 
well. Moreover, gluons can be linearly polarized, even if their parent proton 
is unpolarized, because of their spin-orbit couplings. 

The unpolarized and polarized TMD gluon distributions are so far experimentally
unkown, although several processes have been suggested to measure them. One
possibility would be to look at azimuthal asymmetries for dijet or heavy quark pair production in electron-proton collisions at a future Electron-Ion Collider\cite{ep2jetLO1,ep2jetLO2}. For similar observables in hadronic collisions\cite{Boer:2009nc} TMD factorization is expected to be broken due to the presence of both initial and final state interactions\cite{Rogers:2010dm}. A process for which the problem of factorization breaking is absent is $pp \to \gamma \gamma X$\cite{Qiu:2011ai}. However this reaction suffers from a huge background from $\pi^0$ decays and contaminations from quark-induced channels.

In the following, after providing a formal definition of gluon TMD distributions in terms of QCD operators, we show how they could be probed in Higgs boson and heavy quarkonium production at the LHC. For the latter process, the extraction of gluon distributions should not be hampered if the two quarks that form the bound state are produced in a colorless state already at short distances.

\section{Definition of gluon TMDs for unpolarized hadrons}

Parton correlators describe the hadron $\to$ parton transitions. They are defined on the light front LF, $\xi\cdot n =0$, where $n$ is a four-vector conjugate 
to the hadron momentum $P$. After expanding the gluon momentum as  $p = x\,P + p_\sT + p^- n$, the correlator for an unpolarized proton can be written as\cite{Mulders:2000sh}  
\begin{eqnarray}
\label{GluonCorr}
\Phi_g^{\mu\nu}(x,\bm p_\sT )
& = &  \frac{n_\rho\,n_\sigma}{(p{\cdot}n)^2}
{\int}\frac{d(\xi{\cdot}P)\,d^2\xi_\sT}{(2\pi)^3}\
e^{ip\cdot\xi}\,
\langle P|\,\tr\big[\,F^{\mu\rho}(0)\,
F^{\nu\sigma}(\xi)\,\big]
\,|P \rangle\,\big\rfloor_{\xi\cdot n=0} \nonumber \\
& = &
-\frac{1}{2x}\,\bigg \{g_\sT^{\mu\nu}\,f_1^g (x,\bm p_\sT^2)
-\bigg(\frac{p_\sT^\mu p_\sT^\nu}{M_p^2}\,
{+}\,g_\sT^{\mu\nu}\frac{\bm p_\sT^2}{2M_p^2}\bigg)
\;h_1^{\perp\,g} (x,\bm p_\sT^2) \bigg \} ,
\label{eq:corr}
\end{eqnarray}
where gauge links have been omitted. In Eq.~(\ref{eq:corr}), $F^{\mu\nu}$ is 
the gluon field strength tensor, $g^{\mu\nu}$ is defined as $g^{\mu\nu}_{\sT} = g^{\mu\nu} - P^{\mu}n^{\nu}/P{\cdot}n-n^{\mu}P^{\nu}/P{\cdot}n$, $p_{\sT}^2 = -\bm p_{\sT}^2$ and $M_p$ is the proton mass. The gluon correlator can therefore be parametrized in terms of two independent TMD distributions: the unpolarized one is denoted by $f_1^g(x,\bm{p}_\sT^2)$, while  $h_1^{\perp\,g}(x,\bm{p}_\sT^2)$ is the 
helicity-flip distribution of linearly polarized gluons, which satisfies the model-independent positivity bound\cite{Mulders:2000sh}
\begin{equation}
\frac{\bm p_\sT^2}{2M_p^2}\,|h_1^{\perp g}(x,\bm p_\sT^2)|\le f_1^g(x,\bm p_\sT^2)\,.\label{eq:Bound}
\end{equation}
Being $T$-even, $h_1^{\perp\,g}$ is nonzero even in absence of initial and final 
state interactions. However, as any other TMD distribution, $h_1^{\perp\,g}$ might receive contributions from such interactions, which can render it nonuniversal.  

\section{The Higgs transverse momentum distribution}

\begin{figure*}[t]
\centering
{\includegraphics[width=0.6\textwidth]{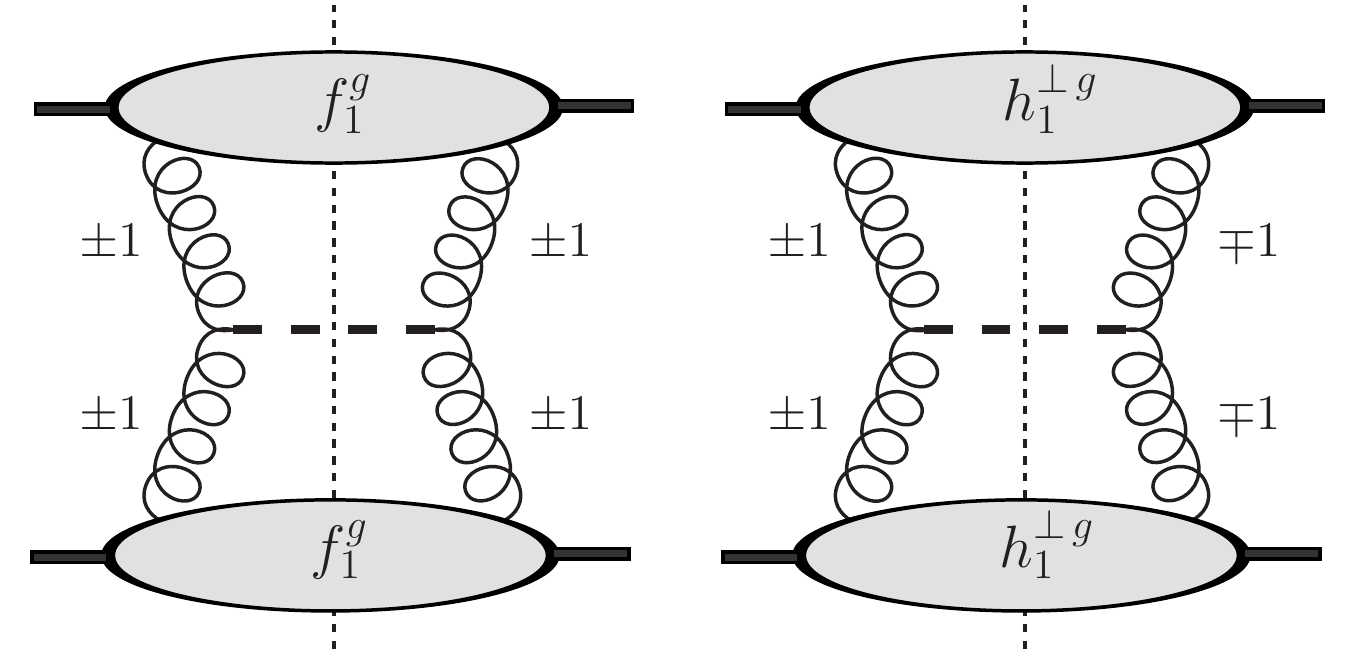}}
\caption{Gluon helicities for the  process $gg\to {\cal S}$, where ${\cal S}$ is
 a (pseudo)scalar particle, for unpolarized (left) and linearly polarized (right) production.}
\label{fig:hel}
\end{figure*}

We consider the inclusive hadroproduction of a Higgs boson, 
\begin{equation}
p(P_A)+p(P_B)\to H(q)+X\,,
\end{equation}
with the four-momenta of the particles given between brackets. Within the framework of TMD factorization, the cross section can be written as
\begin{eqnarray}
\d\sigma
& = &\frac{1}{2 s}\,\frac{d^3 \bm q}{(2\pi)^3\,2 q^0} 
{\int} \d x_a \,\d x_b \,\d^2\bm p_{a\sT} \,\d^2\bm p_{b\sT}\,(2\pi)^4
\delta^4(p_a{+} p_b {-} q)
 \nonumber \\
&&\qquad \times
\,  \Phi_g^{\mu\nu}(x_a {,}\,\bm p_{a \sT})\, \Phi_g^{\rho\sigma}(x_b {,}\,\bm p_{b \sT})\overline{\sum_{\rm colors}}
 {\cal A}_{\mu\rho}\, {\cal A}_{\nu\sigma}^*\,(p_a, p_b; q)\,,
\label{CrossSec}
\end{eqnarray}
where  $s = (P_A + P_B)^2$ denotes the total energy squared in the hadronic center-of-mass frame, while ${\cal A}$ is the hard scattering amplitude for the 
dominant gluon-gluon fusion subprocess $g(p_a)\,+\,g(p_b)\,\to\, {H}(q)$. We 
take all quark masses to be zero, except for the top quark mass $M_t$, and 
neglect electroweak corrections. Therefore the Higgs boson can couple to the 
gluon only via a triangular top quark loop. When the transverse momentum
of the Higgs boson is small, $\vert \bm q_\sT \vert \ll M_H$, and in the limit $M_t\to\infty$, 
the cross section has the following structure\cite{TMDHiggs1,TMDHiggs2}
\begin{eqnarray}
\frac{\d\sigma}{\d y\, \d^2 \bm q_\sT}  =  \frac{\pi}{576}\,\frac{M_H^2}{s\,v^2} \,\left(\frac{\alpha_{s}}{\pi}\right)^{2}\,\left \{ {\cal{C}}\left[f_{1}^{g}\, f_{1}^{g}\right] + {\cal{C}} [w_{0}\, h_{1}^{\perp g}\, h_{1}^{\perp g}]\right \}\,,\label{eq:CSscalarHiggs} 
 \end{eqnarray}
where $y$ is the rapidity of the Higgs boson along the direction of the incoming protons and $v\approx 246$ GeV is the vacuum expectation value of the Higgs field. The two different contributions to the cross section, due to the unpolarized and linearly polarized gluon distributions, are depicted in Fig.~\ref{fig:hel}. The convolution of two TMD distributions $f$ and $g$ is defined as 
\begin{eqnarray}
\mathcal{C}[w\, f\, g]  \equiv  \int d^{2}\bm p_{a\sT}\int d^{2}\bm p_{b\sT}\,
\delta^{2}(\bm p_{a\sT}+\bm p_{b\sT}-\bm q_{\sT})\, w(\bm p_{a\sT},\bm p_{b\sT})\, f(x_{a},\bm p_{a\sT}^{2})\, g(x_{b},\bm p_{b\sT}^{2})\,,\label{eq:Conv}\nonumber \\
\end{eqnarray}
with $x_{a,b} = M_H e^{\pm y}/\sqrt{s}$ and the transverse weight given by
\begin{equation} 
 w_0= \frac{1}{2M_p^{4}} \, \left [ (\bm p_{a\sT}\cdot\bm p_{b\sT})^{2}-\frac{1}{2} \, \bm p_{a\sT}^{2}\bm p_{b\sT}^{2} \right ]\,.
\end{equation}

\begin{figure*}[t]
\centering
\vspace{-2cm}
{\includegraphics[width=0.45\textwidth]{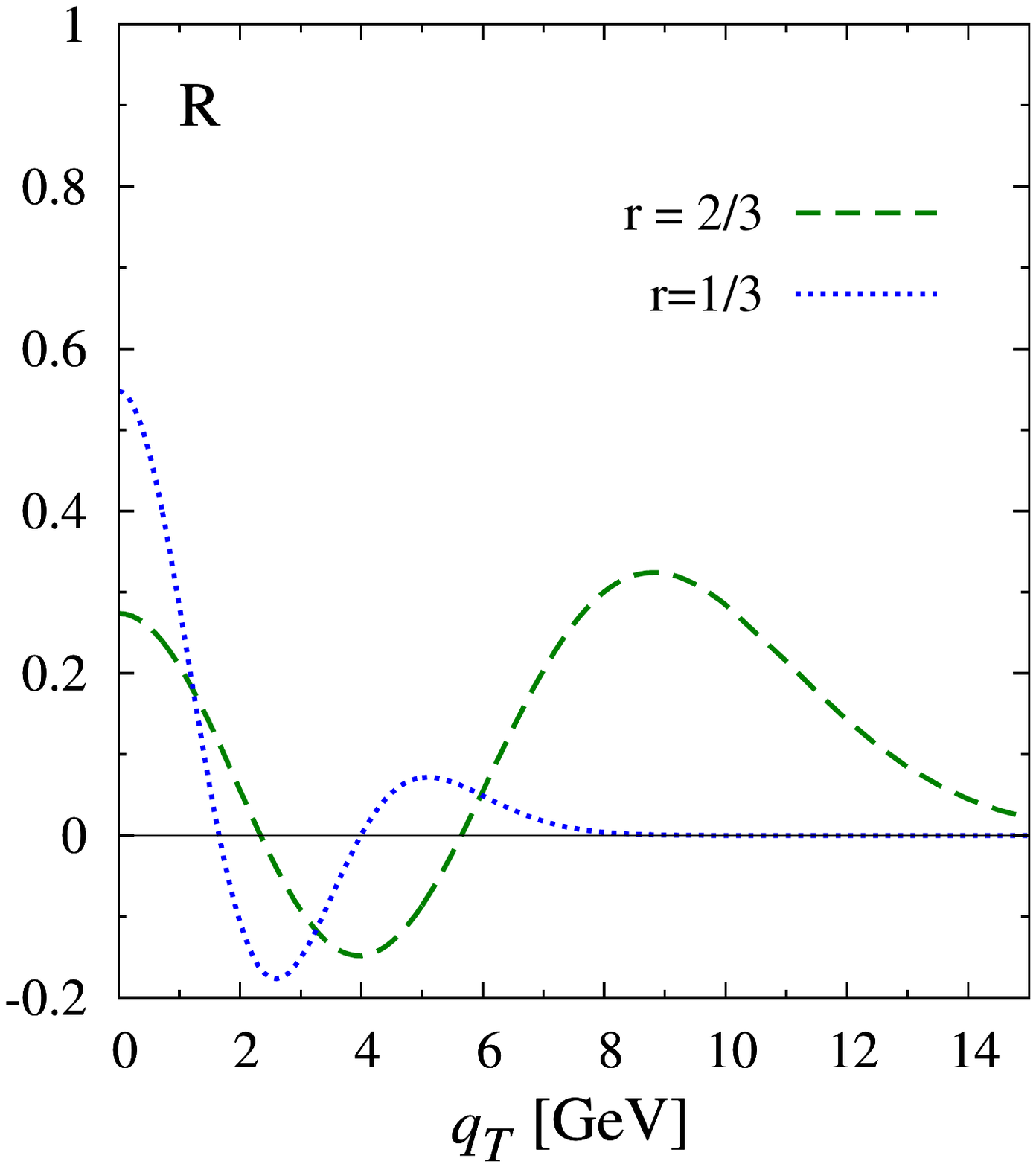}}
{\includegraphics[width=0.45\textwidth]{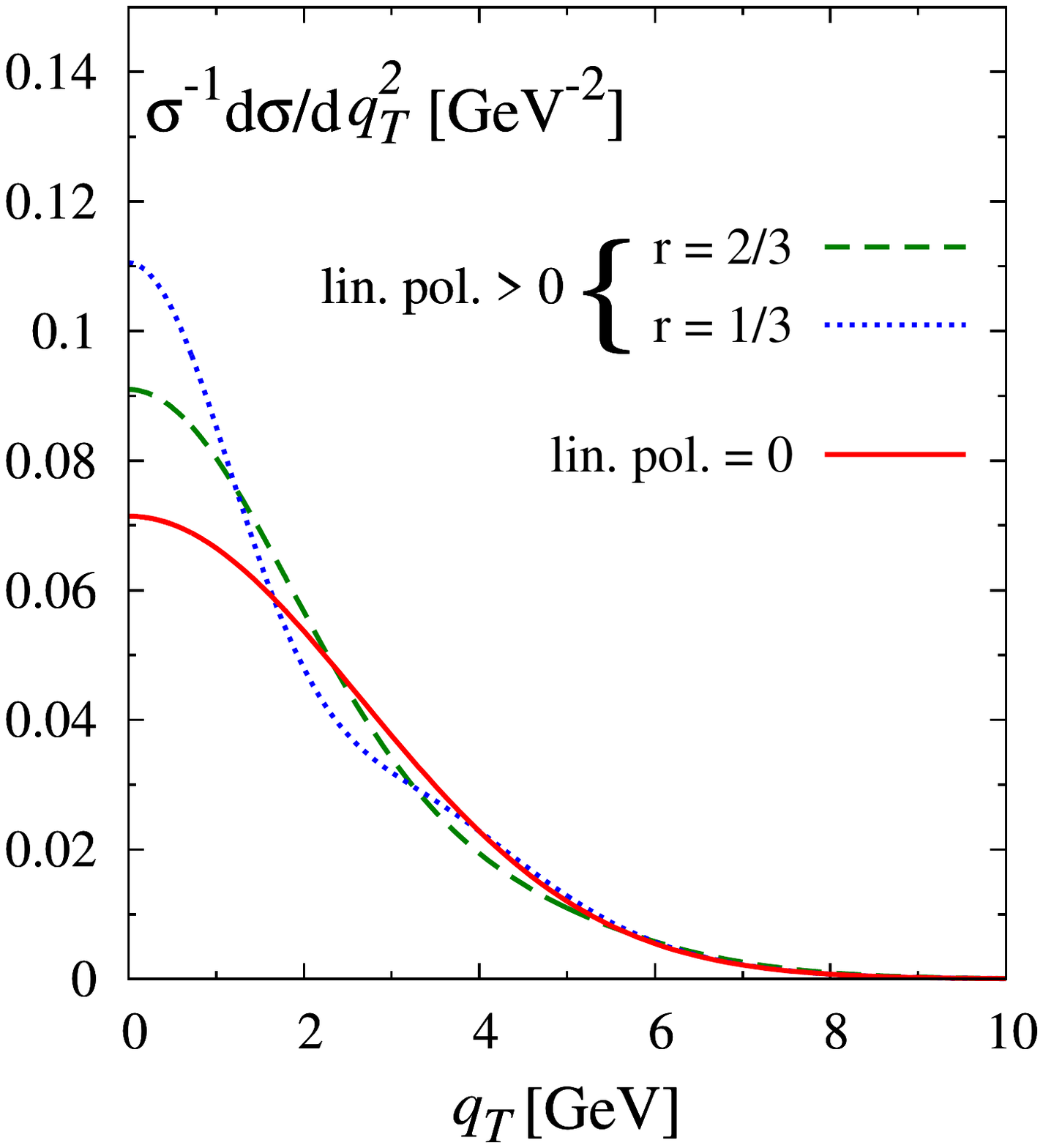}}
\caption{Ratio $R =  {{\cal{C}} [w_0\, h_{1}^{\perp \,g}\, h_{1}^{\perp \,g} ]}/{\mathcal{C} [f_{1}^{g}\, f_{1}^{g} ]}$ (left panel) and transverse momentum distribution for a Higgs boson (right panel) as a function of $q_\sT$ at rapidity $y=0$. Results are obtained adopting the input distributions in Eqs.~(\ref{eq:Gaussf1})-(\ref{eq:Gaussh1perp}) with  $\langle p_\sT^2\rangle = $ 7 GeV$^2$ and considering two different values of $r$: $r=2/3$ and $r=1/3$.}
\label{fig:higgsqT}
\end{figure*}

In the following, we assume that the gluon distributions have a simple Gaussian
 dependence on transverse momentum. Namely,
\begin{equation}
f_1^g(x,\bm p_\sT^2) = \frac{f_1^g(x)}{\pi \langle  p_\sT^2 \rangle}\,
\exp\left(-\frac{\bm p_\sT^2}{\langle  p_\sT^2 \rangle}\right)\,,\label{eq:Gaussf1}
\end{equation}
where $f_1^g(x)$ is the collinear gluon distribution and the width
$\langle p_\sT^2 \rangle$ is taken to be independent  of $x$ and the energy scale set by $M_{H}$, while $h_1^{\perp\,g}$ has the form
\begin{equation}
h_1^{\perp g}(x,\bm p_\sT^2)=\frac{M_p^2 f_1^g(x)}{\pi \langle p_\sT^2 \rangle^2}\frac{2(1-r)}{r}\,
\exp\left(1 -\frac{1}{r}\,\frac{\bm p_\sT^2}{\langle p_\sT^2 \rangle}\right)\,,\label{eq:Gaussh1perp}
\end{equation}
which satisfies the bound in Eq.\ (\ref{eq:Bound}), although it does not 
saturate it everywhere, and $0<r<1$. The resulting transverse momentum distribution of the Higgs boson,
\begin{equation}
\frac{1}{\sigma} \, \frac{\d\sigma}{\d y\, \d \bm q_\sT^2} =   
\frac{{\cal C} [f_1^g f_1^g]}{\int \d\bm q_\sT^2 \, {\cal C} [f_1^gf_1^g] } \,\left [1+R(\bm q_\sT^2)\right ] \, ,
\end{equation}
with $\sigma = \int_0^{\infty} \d \qT^2\, \d \sigma$ and
\begin{equation}
R (\bm q_{\sT}^2) = \frac{{\cal{C}}[w_0\, h_{1}^{\perp \,g}\, h_{1}^{\perp \,g} ]}{{\cal{C}}[f_{1}^{g}\, f_{1}^{g} ]}\,,
\end{equation}
is presented in Fig.~\ref{fig:higgsqT} (right panel) for $\langle p_{\sT}^2\rangle = 7$ GeV$^2$ and two different values of the parameter $r$. The ratio 
$R$ measures the relative size of the contribution by linearly polarized gluons
and it is shown separately in the left panel of the figure. Its double node
structure is a characteristic feature of the Gaussian model adopted.

\begin{figure*}[t]
\centering
\vspace{-2cm}
{\includegraphics[width=0.45\textwidth]{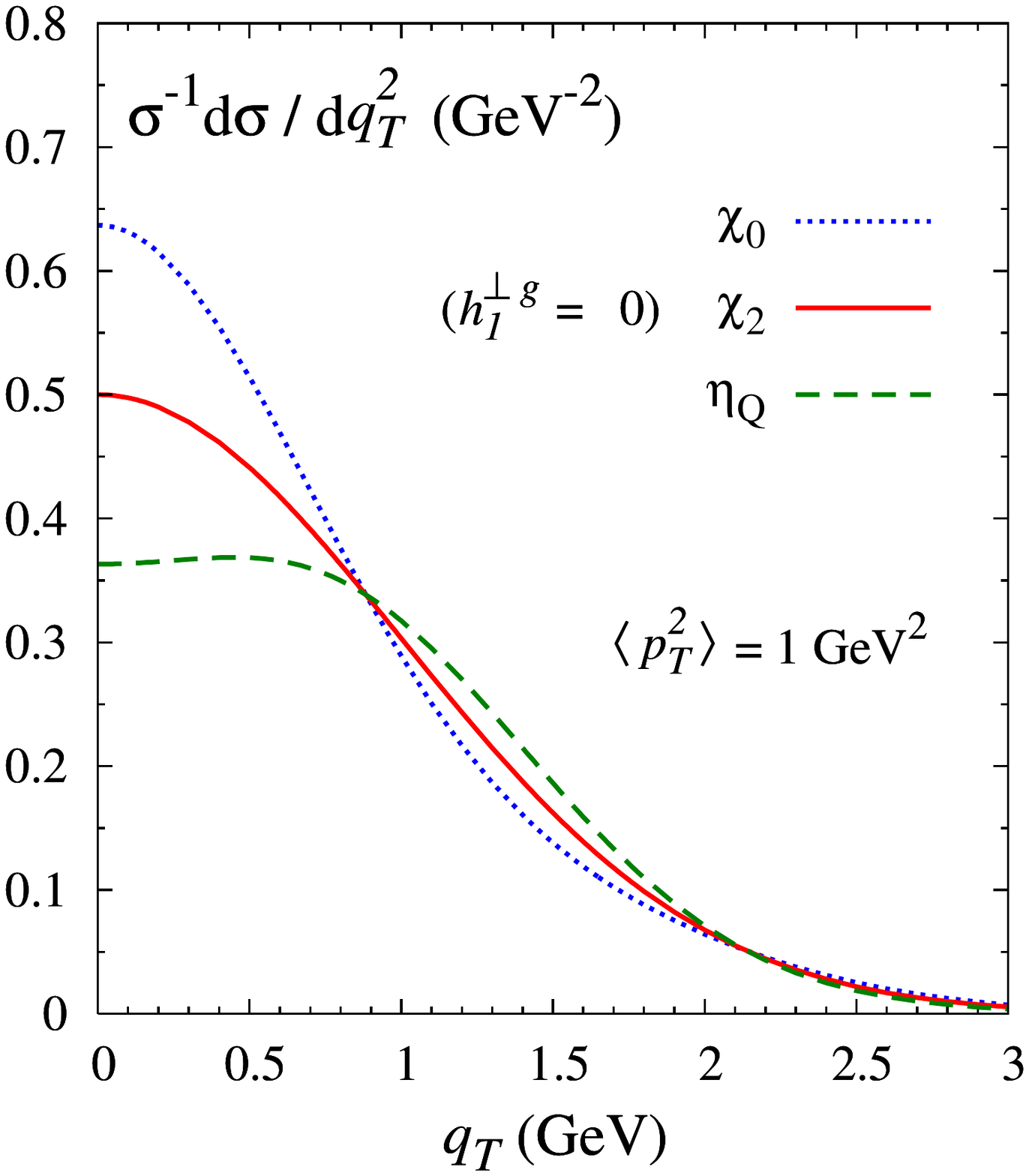}}
{\includegraphics[width=0.45\textwidth]{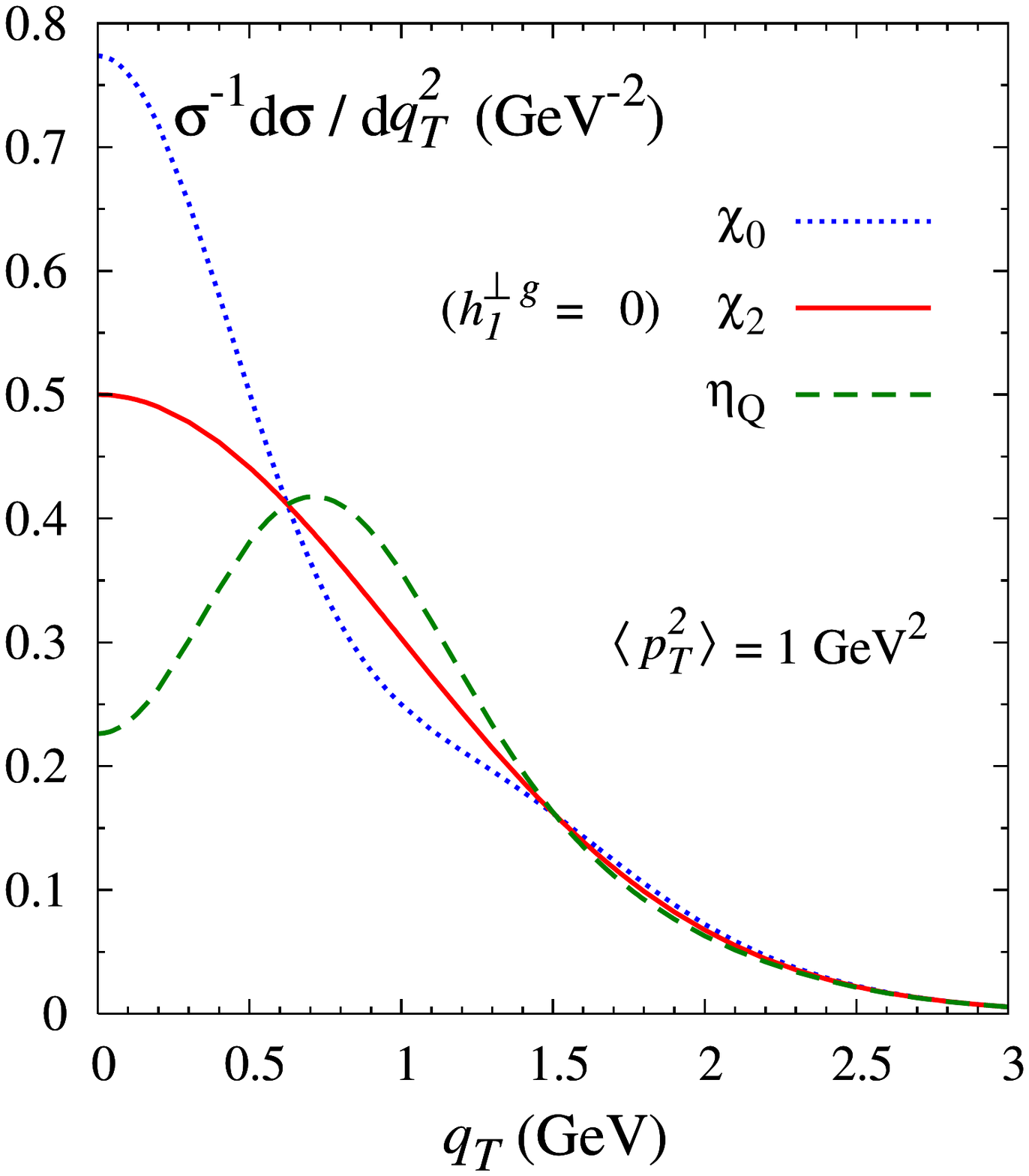}}
\caption{Transverse momentum distributions for different $C=+$ quarkonia as a function of $q_\sT$ at rapidity $y=0$. Results are obtained using the input distributions in Eqs.~(\ref{eq:Gaussf1})-(\ref{eq:Gaussh1perp}), with $\langle p_\sT^2 \rangle = 1$ GeV$^2$ and two different values of $r$: $r=2/3$ (left) and $r=1/3$ (right).}
\label{fig:dsigmadqT}
\end{figure*}

\section{Transverse momentum distributions of $C=+$ quarkonia}

The calculation of the cross section for the process 
\begin{equation}
p(P_A)+p(P_B)\to {\cal Q}(q)+X\,, 
\end{equation}
where ${\cal Q}$ is a heavy quark-antiquark bound state with even charge conjugation, $C=+$, is very similar to the one outlined in the previous section for inclusive Higgs production.  The amplitude ${\cal A}$ in  Eq.~(\ref{CrossSec}) now
refers to the dominant subprocess $g(p_a)\,+\,g(p_b)\,\to\, {\cal{Q}}(q)$ and 
is evaluated at order $\alpha_s^2$ within the framework of the color-singlet model. Color octet contributions are expected to  be negligible, according to nonrelativistic QCD arguments\cite{ppJPsiLO}. For small transverse momentum, $\qT^2 \ll M^2_{\cal Q}$, with $M_{\cal Q}$ being the quarkonium mass, we find that the
transverse momentum distributions for $\eta_Q$ and $\chi_{Q0,2}$ ($Q=c$, $b$)  
are  given by
\begin{eqnarray}
\frac{1}{\sigma(\eta_Q)} \, \frac{\d\sigma (\eta_Q)}{\d y\, \d \bm q_\sT^2} & = &  
\frac{{\cal C} [f_1^g f_1^g]}{\int \d\bm q_\sT^2 \, {\cal C} [f_1^gf_1^g] } \,\left [1-R(\bm q_\sT^2)\right ] \, ,\nonumber \\
\frac{1}{\sigma(\chi_Q)} \, \frac{\d\sigma (\chi_{Q 0})}{\d y \,\d \bm q_\sT^2}  & = &  \frac{{\cal C} [f_1^g f_1^g]}{\int \d\bm q_\sT^2 \, {\cal C} [f_1^gf_1^g] }\, \left [1+R(\bm q_\sT^2)\right ] \, ,\nonumber \\
\frac{1}{\sigma(\chi_Q)} \, \frac{\d\sigma (\chi_{Q 2})}{\d y\, \d \bm q_\sT^2}  & = & \frac{{\cal C} [f_1^g f_1^g]}{\int \d\bm q_\sT^2 \, {\cal C} [f_1^gf_1^g] }\,.
\end{eqnarray}
It turns out that the cross sections for scalar and pseudoscalar quarkonia are
modified in different ways by linearly polarized gluons. In particular, 
the distributions for $\chi_{c,b\,0}$ are similar to the one for a (scalar) 
Higgs boson discussed above. Polarization effects 
contribute with  a negative sign to the distributions of  $\eta_{c,b}$, while 
they are strongly suppressed for higher angular momentum quarkonium states.
Our numerical estimates are presented in Fig.~\ref{fig:dsigmadqT} and show that,  in principle, the $q_\sT$-distributions for (pseudo)scalar quarkonia could be used to probe $h_1^{\perp\,g}$, while $f_1^g$ can be accessed by looking at $\chi_{c,b\,2}$. Moreover, a comparison among the different spectra could help to cancel out uncertainties. 

Since particles resulting from  a $2\to 1$ scattering 
process typically have a small transverse momentum and are mostly lost along the beam pipe at collider facilities like the LHC, forward detectors like the LHCb
are required. In this respect, the first measurement of inclusive $\eta_c$ hadroproduction\cite{Aaij:2014bga}, via its $p\bar p$ decay channel, is very encouraging. Finally, from the theoretical point of view, we point out that TMD factorization has recently been estabilished at next-to-leading order for the process $p\,p\to \eta_{c,b}\, X$\cite{Ma:2012hh}, but not for $p\,p\to  \chi_{c,b}\, X$\cite{Ma:2014oha}.

\section{$C=-$ quarkonium production in association with a photon}

\begin{figure}[t]
\centering
\includegraphics[width=0.7\columnwidth]{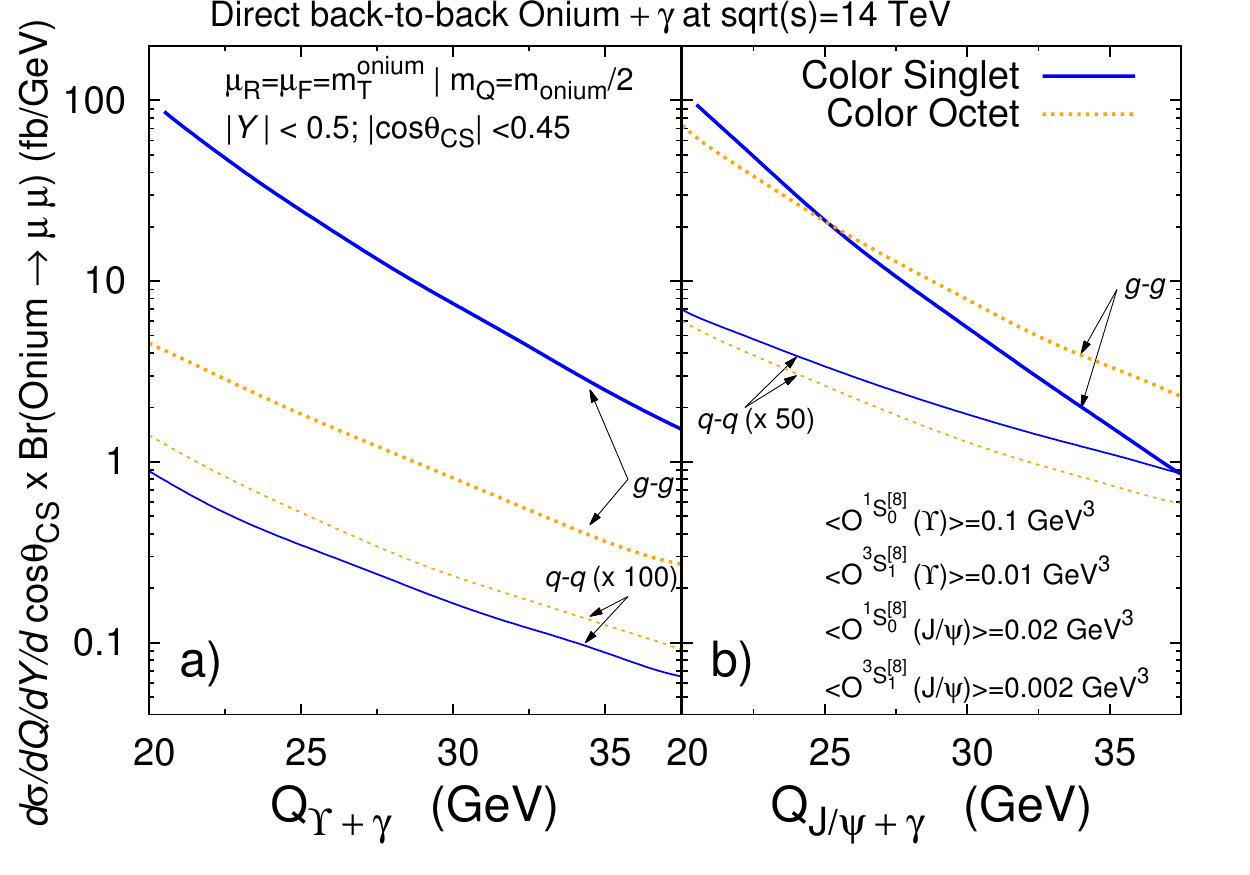}
\caption{Color-octet and color-singlet contributions to the production of a photon with a) an $\Upsilon(1S)$ and  b) a $J/\psi$  from $g-g$ fusion and $q-\bar q$ annihilation channels as a function of the invariant mass of the pair $Q$. The curves for the $q-\bar q$ annihilation are rescaled by a factor 100 in a) and 50 in b). }
\label{fig:prod_channels}
\end{figure}

Polarized and unpolarized TMD gluon distributions could be also accessed
through the reaction\cite{Qgamma}
\begin{equation}
p(P_A)+p(P_B)\to {\cal Q}(P_{\cal Q}) + \gamma (P_{\gamma})+ X\,,
\end{equation}
 where now ${\cal Q}$ is a $C=-$ quarkonium ($J/\psi$ or $\Upsilon$) produced almost back to back with a photon. The momentum imbalance of the pair in the final state, $\qT = \bm P_{{\cal Q}\sT}+\bm P_{\gamma\sT}$, is small, but not the individual transverse momenta of the two particles. Hence no forward detector is needed in this case. In this particular configuration, the process is expected to be
dominated by color-singlet contributions\cite{Mathews:1999ye}, hence TMD 
factorization should be applicabile.

We find that the resulting cross section can be written in the following 
form,
\begin{eqnarray}\label{eq:crosssection}
\frac{\d\sigma}{\d Q \d Y \d^2 \qT \d \Omega} 
 & = & \frac{4\alpha_s^2\alpha^2e^2_Q\vert R_0(0)\vert^2}{3M_{\cal Q}^2}\, \frac{Q^2-M_{\cal Q}^2 }{s Q^3 D} \,\left\{
  F_1\, \mc{C} \Big[f_1^gf_1^g\Big]\right . \nonumber \\
& & \left . + F_3 \,\mc{C} \Big[w_3 f_1^g h_1^{\perp g} + (x_a\! \leftrightarrow\! x_b )\Big] \cos 2\phi + \,F_4  \,\mc{C} \left[w_4 h_1^{\perp g}h_1^{\perp g}\right]\cos 4\phi  \right \}\,,\nonumber \\
\label{eq:Qgamma}
\end{eqnarray}
with  $Q$ and $Y$ being, respectively, the invariant mass and the rapidity of the pair. Similarly to $\bm q_{\sT}$, these quantities are measured in the hadronic center-of-mass frame, while the solid angle $\Omega=(\theta,\phi)$ is measured in the Collins-Soper frame\cite{Collins:1977iv}. This is defined as the frame in which the final pair is at rest, with the $\hat x\hat z$-plane spanned by 
$(\bm P_A$, $\bm P_B)$ and the $\hat x$-axis set by their bisector.   
In Eq.~(\ref{eq:Qgamma}), $R_0(0)$ is the quarkonium radial wave function calculated at the origin and $e_Q$ is the heavy quark charge in units of the proton 
charge. The factors $F_i$ and the denominator $D$ are given by  
\begin{eqnarray}
 F_1 & = &  1 + 2 \gamma ^2 + 9 \gamma ^4 + (6 \gamma ^4-2) \cos^2\theta + (\gamma ^2-1)^2 \cos^4\theta\,, \nonumber\\
 F_3 &= &  4\, \gamma^2\, \sin^2\theta\,, \\
 F_4 & =&  (\gamma ^2-1)^2 \sin^4\theta  ,\nonumber\\
 D &= &  \left[(\gamma ^2+1)^2-(\gamma^2-1)^2 \cos^2 \theta\right]^2\,,
\end{eqnarray}
with $\gamma \equiv Q/M_\mc{Q}$. The explicit expressions for the 
transverse weights are  
\begin{equation}
 w_3 = \frac{\qT^2\bm p_{b\sT}^2 - 2 (\qT{\cdot}\bm p_{b \sT})^2}{2 M_p^2 \qT^2},
\quad
 w_4 =  2\left[\frac{\bm p_{a\sT}{\cdot}\bm p_{b\sT}}{2M_p^2} - 
		\frac{(\bm p_{a\sT}{\cdot}\qT) (\bm p_{b\sT}{\cdot}\qT)}{M_p^2\qT^2}\right]^2 -\frac{\bm p_{a\sT}^2 \bm p_{b \sT}^2 }{4 M_p^4}\,,
\end{equation}
and the light-cone momentum fractions are $x_{a,b} =  \exp[\pm Y]\, Q/\sqrt{s}$.

By measuring the following three observables,
\begin{equation}
{\cal S}^{(n)}_{q_T} \equiv  \frac{\int \d\phi\, \cos(n\, \phi )\, \frac{\d\sigma}{\d Q \d Y \d^2 \qT \d \Omega}}
{\int \d \bm q_\sT^2 \int \d\phi \,\frac{\d\sigma}{\d Q \d Y \d^2 \qT \d \Omega}}\,,
\end{equation}
with  $n=0,2,4$, and the $q_\sT^2$ integration in the denominator up to $Q^2/4$, we are able to single out the three terms in Eq.~(\ref{eq:Qgamma}). We obtain
\begin{eqnarray}
\label{eq:qTdistrs}
{\cal S}^{(0)}_{q_T} & = &\frac{\mc{C}[f_1^g f_1^g]}
  {\int  \d \bm q_\sT^2\, \mc{C}[f_1^g f_1^g]},\nonumber \\
  {\cal S}^{(2)}_{q_T} & = & 
  \frac{F_3\, \mc{C}[w_3 f_1^g h_1^{\perp g} + (x_a \leftrightarrow x_b)]}
  {2 F_1 \int  \d \bm q_\sT^2\, \mc{C}[f_1^g f_1^g]},\nonumber\\ 
{\cal S}^{(4)}_{q_T} &= & 
 \frac{F_4\, \mc{C}[w_4 h_1^{\perp g} h_1^{\perp g}]}
  {2 F_1 \int \d \bm q_\sT^2\, \mc{C}[f_1^g f_1^g]}\,.
\label{eq:S}
\end{eqnarray}

\begin{figure*}[t]
\centering
{\includegraphics[width=0.31\textwidth]{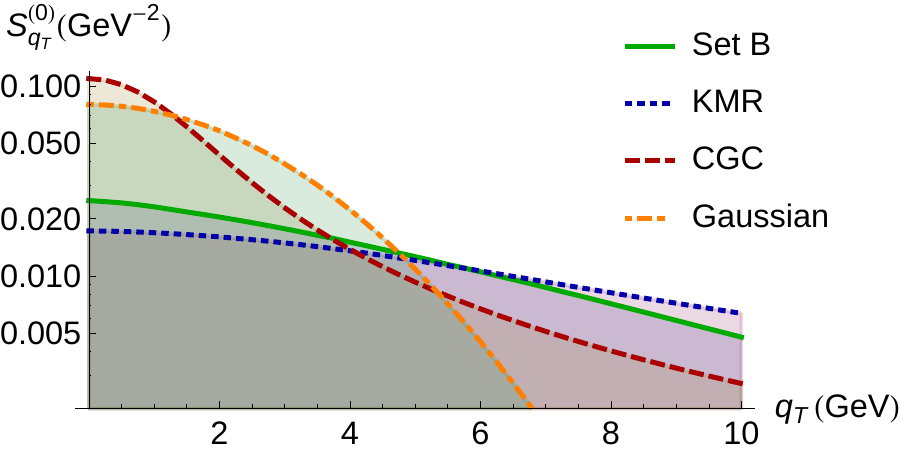}}
\hspace{0.2cm}
{\includegraphics[width=0.31\textwidth]{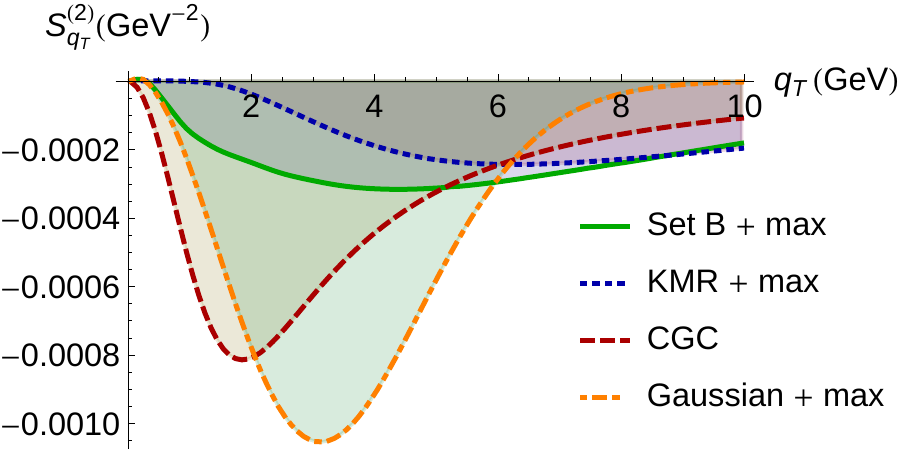}}
\hspace{0.2cm}
{\includegraphics[width=0.31\textwidth]{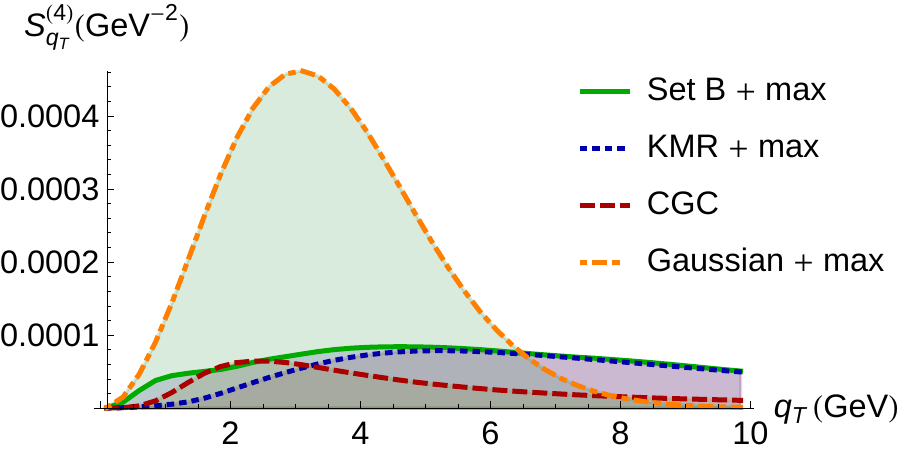}}
\caption{Estimates for the observables ${\cal S}^{(0)}_{q_T}$,  ${\cal S}^{(2)}_{q_T}$ and  ${\cal S}^{(4)}_{q_T}$ given in Eq.~(\ref{eq:S}) for the process $p\,p\to {\cal Q}\, \gamma\,  X$ as a function of the momentum imbalance of the pair, 
$q_\sT$, in the kinematic region defined by  $\sqrt{s}=14$ TeV, $Q=20$ GeV, $Y=0$, $\theta =\pi/2$, $x_a=x_b \simeq1.4\times10^{-3}$.}
\label{fig:dsigma4dqT}
\end{figure*}

We provide numerical estimates for $\Upsilon+\gamma$ production in a kinematic region where color octect contributions are suppressed\cite{Qgamma}, as illustrated in Fig.~\ref{fig:prod_channels}.  Our results, obtained adopting different models for the TMD distributions, are presented in Fig.~\ref{fig:dsigma4dqT}. It turns out that the size of ${\cal S}^{(0)}_{q_T}$ should be sufficient to allow for an extraction of $f_1^g$ as a function of $q_\sT$. On the contrary, ${\cal S}^{(2)}_{q_T}$ and ${\cal S}^{(4)}_{q_T}$ are quite small and one would need to integrate them over $\bm q_\sT^2$, {\it i.e.}~up to $Q^2/4$, to have an experimental evidence of a nonzero $h_1^{\perp\,g}$.

\section{Conclusions}

Within a TMD factorization approach, we have calculated the transverse momentum
distributions for the inclusive hadroproduction of Higgs bosons and $C=+$ quarkonia. We find that the linear polarizazion of gluons inside unpolarized protons
modifies these observables in a very characteristic way, leading to different modulations of the transverse spectra of scalar ($H$, $\chi_{c0}$, $\chi_{b0}$) and pseudoscalar ($\eta_c$, $\eta_b$) particles,  depending on their parity.
 On the contrary, the distributions of $\chi_{c2}$ and $\chi_{b2}$ are not affected. Such measurements do not require any angular analysis and could be performed by the LHCb Collaboration or at the proposed fixed-target experiment AFTER@LHC\cite{after1,after2}. Moreover, a first determination of the polarized and unpolarized TMD gluon distributions could come from the study of, respectively, azimuthal asymmetries and transverse spectra in $p\,p \to J/\psi(\Upsilon) \, \gamma\, X$  at the LHC. We have checked that yields are large enough to perform these analyses using already existing data at the center-of-mass energies $\sqrt{s}=7$ and $8$ TeV\cite{Qgamma}.  

\section*{Acknowledgments}
I would like to thank Dani\"el Boer,  Stan Brodsky, Maarten Buffing, Wilco den Dunnen, Jean-Philippe Lansberg, Piet Mulders, Marc Schlegel and Werner Vogelsang, with whom I have collaborated on this topic over the past few years. This work was supported by the European Community under the “Ideas” program QWORK (contract 320389).

\end{document}